\documentclass[12pt,a4paper,prd,onecolumn,superscriptaddress,floatfix,nofootinbib,showpacs]{revtex4}

\usepackage[margin=2.0cm]{geometry}
\usepackage[utf8]{inputenc}
\usepackage[english]{babel}
\usepackage{graphicx}
\usepackage{color}
\usepackage{float}
\usepackage{amsmath}
\usepackage{amssymb}
\usepackage{hyperref}
\usepackage{natbib}
\usepackage{journalnew}

\newcommand{\dl}{\ensuremath{D_{\rm L}}}
\newcommand{\da}{\ensuremath{D_{\rm A}}}

\newcommand{\rmd}{\mathrm{d}}
\newcommand{\rmi}{\mathrm{i}}
\begin{document}

\title{Modifications to the Etherington Distance Duality Relation and Observational Limits}
\author{Surhud More}\email{surhud.more@ipmu.jp}
\affiliation{Kavli Institute for the Physics and Mathematics of the Universe (WPI), Todai Institutes of Advanced Study, University of Tokyo, 5-1-5 Kashiwanoha, Kashiwa 277-8583, Japan}
\author{Hiroko Niikura}
\affiliation{Kavli Institute for the Physics and Mathematics of the Universe (WPI), Todai Institutes of Advanced Study, University of Tokyo, 5-1-5 Kashiwanoha, Kashiwa 277-8583, Japan}
\author{Jonas Schneider}
\affiliation{Department of Physics, Institute for Quantum Gravity, Friedrich-Alexander University Erlangen-Nuremberg, Staudtstrasse 7, 91058 Erlangen, Germany}
\author{Frederic P. Schuller}
\affiliation{Department of Physics, Institute for Quantum Gravity, Friedrich-Alexander University Erlangen-Nuremberg, Staudtstrasse 7, 91058 Erlangen, Germany}
\author{Marcus C. Werner}
\affiliation{Yukawa Institute for Theoretical Physics and Hakubi Center for Advanced Research, Kyoto University, Kitashirakawa Oiwakecho Sakyoku, Kyoto 606-8502, Japan}
\date{\today}
\begin{abstract}
The Etherington distance duality relation, which relates the luminosity distance, the angular diameter distance and the redshift of objects, depends only upon a conservation law for light that traces back directly to the Lorentzian spacetime geometry. We show that this duality relation indeed survives transition to the most general linear electrodynamics without birefringence, which rests on a spacetime geometry constituted by a Lorentzian metric and two scalar fields, a dilaton and an axion. By computing the Poynting vector and optical scalar transport in the geometrical optics limit of this framework, we derive the modification of the light flux in the presence of a dilaton field and present improved constraints on the gradient of the dilaton field from observations of the Cosmic Microwave Background spectrum. Although this flux modification may seem applicable also to fundamental modifications of the Etherington relation, we show that the distance duality relation still holds true. Thus any deviations within this classical theory would imply non-metricities, once astrophysical sources for attenuation, such as dust, are accounted for. Moreover, using the most up-to-date measurements of the luminosity distances to Supernovae of Type Ia, and the inferred angular diameter distances from the baryon acoustic feature measurements, we perform a scale-free and nearly model-independent test of the Etherington distance duality relation between redshifts of $0.38$ and $0.61$. We find consistency with the standard distance duality relation and constrain the optical depth of light between these two redshifts to $\Delta\tau=-0.006\pm0.046$.
\end{abstract}

\maketitle

\section{Introduction}
The Etherington distance duality relation, whose original version was derived in 1933 \cite{Etherington:1933}, establishes a connection between the luminosity distance $\dl$, redshift $z$ and angular diameter distance $\da$ (e.g., \cite{Schneider:1992}, p.116),
\begin{equation}
\dl = (1+z)^2 \da,
\label{eth}
\end{equation}
which is a purely kinematical result in the sense that it depends only on (i) a conservation law for light rays in geometrical optics, and (ii) the Lorentzian spacetime geometry. It is thus independent of the dynamics of the background geometry described by gravitational field equations, such as those of general relativity. Therefore, observational testing of the Etherington relation amounts to testing the fundamental kinematical structure of spacetime. 
\\
Such tests have now become possible due to the observational advances in recent years. \citet{Bassett:2004a} proposed and carried out a test of the Etherington relation using luminosity distances from SNeIa and angular diameter distances estimated from FRIIb radio galaxies, compact radio sources and X-ray clusters, in order to constrain the transparency of the Universe \citep[see also][]{Uzan:2004, DeBernardis:2006, Jackson:2008, Avgoustidis:2009, Holanda:2011, Liang:2011, Nair:2011, Khedekar:2011, Cao:2011, Liao:2013, Shafieloo:2013, Liang:2013, Nair:2015, Liao:2015, Rana:2016}. The angular diameter distance measurements, however, may suffer from a number of systematic uncertainties in the estimates of the sizes of these astrophysical objects. In their stead, \citet{More:2009} suggested the use of the ratio of the angular diameter distance $\da$ measured using the baryon acoustic feature (BAF) as a standard ruler, and the $\dl$ measured using type Ia supernovae (SNeIa) as standardizable candles, at two different redshifts to provide a robust test of the Etherington relation, and presented constraints on cosmic opacity using then available data \citep[see][for further follow-up work on this topic]{Nair:2012, Chen:2012, Wu:2015}.

The tests of the Etherington relation so far have focused on constraining the absorption of light (rays) by introducing an effective opacity $\tau$ between the observer (at $z=0$) and a SNeIa at redshift $z$ such that the observed luminosity distance $D_{\rm L, obs}$ is related to the geometrical $\dl$ by
\[
D_{\rm L, obs}^2(z)= \dl^2(z)e^{\tau(z)}.  
\]
While $\tau(z)>0$ would naturally occur due to absorption, detection of a negative $\tau$ would suggest a fundamental effect in the kinematical structure of spacetime. A number of studies have also assumed various phenomenological functional forms to characterize the deviation from the Etherington relation \citep[e.g., see][]{Avgoustidis:2009, Holanda:2011}. In light of this, it is interesting to study the sources of possible modifications of the Etherington distance duality relation from a theoretical point of view. In this paper, we investigate how such modifications may arise from a classical (geometrical optics) point of view, starting with a conservative approach: 
Insisting that there be no violation of local Lorentz symmetry, the maximally possible linear extension of electrodynamics consists in introducing both a dilaton and an axion field. In section \ref{sec-th}, we show, however, that this does not modify the Etherington relation. For negative absorption this would directly hint at a more refined spacetime geometry than the one provided by a metric, while for positive absorption, mundane astrophysical sources such as dust would have to be ruled out conclusively before such conclusions on the spacetime structure itself would be justified. New observational constraints on such deviations obtained from the latest BOSS survey data and the recently updated SNeIa collection sample are then presented in section \ref{sec-obs}. Future prospects and the application of non-metric theories of gravity to this problem are discussed in section \ref{sec-con}. Throughout this paper, we employ metric signature $(+,-,-,-)$, Latin indices for spacetime coordinates and use units such that $c=1$.

\section{Theoretical viability of modifications}
\label{sec-th}
\subsection{Vacuum electrodynamics}
We begin by considering possible modifications of vacuum electrodynamics. Assuming that we have an action principle and linear electrodynamics, such that the superposition principle for light holds, then the most general electrodynamics is given in the premetric framework (e.g., \cite{Post:1962, Rubilar:2002, Hehl:2002}), which does not presuppose an underlying Lorentzian spacetime metric, by
\begin{equation}
\mathcal{S}[A]=\int \rmd^4 x \ \mathcal{L}=-\frac{1}{8}\int \rmd^4 x \ \chi^{abcd}F_{ab}F_{cd},
\label{em}
\end{equation}
where the electromagnetic field strength tensor is $F_{ab}=2\partial_{[a}A_{b]}$ as usual, and $\chi^{abcd}$ is the constitutive tensor density with symmetries
\[
\chi^{abcd}=-\chi^{bacd}=-\chi^{abdc}, \ \ \chi^{abcd}=\chi^{cdab},
\]
which characterizes the vacuum of such refined electrodynamics. Inside a dielectric medium, e.g., a crystal, such a premetric structure can be used to effectively describe effects routinely seen in the laboratory. In particular, premetric electrodynamics allow for a birefringence of light, i.e, the splitting of one non-polarized light ray into two of definite polarization. However, if we insist upon the absence of birefringence so that the path of light rays is still governed by a single Lorentzian metric $g_{ab}$ independent of polarization, then the most general constitutive tensor density may be written \cite{Lammerzahl:2004}
\begin{equation}
\chi^{abcd}= \sqrt{-g}\left(g^{ac}g^{bd}-g^{ad}g^{bc}\right)\psi +\phi \epsilon^{abcd},
\label{chi}
\end{equation}
where $\sqrt{-g}=\sqrt{-\det g_{ab}}$ for short, and $\psi$ and $\phi$ are scalar fields called dilaton and axion, respectively. This framework seems particularly suited to fundamental classical modifications of Etherington distance duality at first glance, since it was shown by Ni \cite{Ni:2014} that the dilaton field can yield an amplification of light which may in turn be constrained by observations of the Cosmic Microwave Background blackbody spectrum (for a detailed study of modifications of the reciprocity relation and greybody effects, see also \cite{Ellis:2013}). While Ni's calculation uses a perturbation about Minkowski spacetime and homogeneous fields, we obtain an improved result in the following, by directly computing the Poynting vector and optical scalar transport.

\subsection{Poynting vector}
By definition of $F_{ab}$ and variation of the electrodynamical action (\ref{em}), one obtains the correspondingly refined electromagnetic field equations 
\begin{eqnarray}
\partial_c\left(\chi^{abcd}F_{ab}\right)&=&0, \label{eom1}\\
\partial_{[a}F_{bc]}&=&0. \label{eom2}
\end{eqnarray}
One can also compute the corresponding source tensor density,
\begin{equation}
T_{ab}=\frac{\delta \mathcal{S}}{\delta g^{ab}}=\frac{\partial \mathcal{L}}{\partial g^{ab}},
\label{t-def}
\end{equation}
which is connected to the physical, or Gotay-Marsden \cite{Gotay:1997,Gotay:2004}, energy-momentum tensor density by
\begin{equation}
T^i{}_j=-2g^{ib} \delta^a{}_j T_{ab},
\label{t-gm}
\end{equation}
whose numerical forefactor is usually absorbed into the definition (\ref{t-def}). Here, however, we wish to keep the conceptual distinction between the source tensor density and the energy-momentum tensor density, in view of the conservation law to be discussed in the next subsection. Then from (\ref{em}) with (\ref{chi}) and (\ref{t-gm}), one finds
\begin{equation}
T^i{}_j=\psi \sqrt{-g}\left(F_{ac}F_{jb}g^{ic}g^{ab}+\frac{1}{4}\delta^i{}_jF_{ab}F^{ab}\right).
\label{t-f}
\end{equation}
Next, we perform a WKB analysis analogous to \cite{Schneider:1992}, pp. 93f and 97f, introducing an eikonal scalar phase function $\mathrm{S}$ defining a wave covector field $k_a=-\partial_a \mathrm{S}$, and
\[
F_{ab}=\mathrm{Re}\left[2k_{[a}A_{b]} \exp\left(\rmi \frac{\mathrm{S}}{\epsilon} \right) \right]=k_{[a}A_{b]}\exp\left(\rmi \frac{\mathrm{S}}{\epsilon} \right)+k^{}_{[a}A_{b]}^\ast \exp\left(-\rmi \frac{\mathrm{S}}{\epsilon} \right).
\]
Then the equations of motion (\ref{eom1}), (\ref{eom2}) imply the null condition $g^{ab}k_ak_b=0$, the transversality condition $g^{ab}A_ak_b=0$, and the geodesic equation $k^bk^a{}_{;b}=0$ with respect to $g_{ab}$, as in the standard vacuum with $\psi \equiv1$, $\phi \equiv 0$. 
\\
Now by expressing the complex amplitude $A_a$ in terms of the scalar amplitude $A$ and the spacelike complex polarization $P_a$ such that
\[
A_a=A P_a \ \ \mbox{and} \ \ A=\sqrt{-g^{ab}A^{}_aA^\ast_b} \ \ \mbox{with} \ \ g^{ab}P^{}_{a}P^\ast_b=-1,
\]
and defining the optical scalar
\[
\theta=\frac{1}{2}k^a{}_{;a}
\]
as usual, we find the transport equation for $A$ from (\ref{eom1}), (\ref{eom2}),
\begin{equation}
\dot{A}=k^a\partial_a A=-A \theta+\frac{1}{2}Ak^a\frac{\partial_a \psi}{\psi}=-A\theta+\frac{1}{2}A\frac{\dot{\psi}}{\psi},
\label{a1}
\end{equation}
which depends on the dilaton field $\psi$, but is independent of the axion field $\phi$. Integrating (\ref{a1}) along a light ray parametrized by $t$ yields
\begin{equation}
A(t)=A(0)\sqrt{\frac{\psi(t)}{\psi(0)}}\exp \left( -\int_0^t \rmd t \ \theta \right).
\label{a2}
\end{equation}
Moreover, the on-shell period-averaged energy-momentum density can be computed from (\ref{t-f}), to obtain
\begin{equation}
\langle T \rangle^i{}_j=-\psi \sqrt{-g}g^{ic}g^{ab}\left(k^{}_{[c}A^{}_{a]}k^{}_{[j}A^\ast_{b]}+k^{}_{[c}A^\ast _{a]}k^{}_{[j}A^{}_{b]}\right) =\frac{1}{2}\psi \sqrt{-g}A^2 k^ik_j.
\label{t-av}
\end{equation}
Thus, we indeed recover Ni's \cite{Ni:2014} amplification factor without spectral distortion depending only on the dilation field and being proportional to $\sqrt{\psi}$, albeit without resorting to a perturbation about Minkowski spacetime and homogeneous fields. However, a comparison with observed fluxes should employ the Poynting vector rather than just the scalar amplitude squared. Taking, for instance, $k^a=(1,1,0,0)$, the Poynting vector is
\begin{equation}
S=\langle T \rangle^1{}_0=\frac{1}{2}\psi \sqrt{-g} A^2
\label{s}
\end{equation}
using (\ref{t-av}), which introduces an additional factor proportional to $\psi$ such that $S\propto \psi^2$. Following Ni's argument to obtain a constraint on the dilation field gradient from the electromagnetic spectrum of the CMB, which is very nearly a blackbody spectrum of temperature $T$ with $S\propto T^4$, we have
\[
 \frac{\delta S}{S}=4\frac{\delta T}{T}= 2\frac{\delta \psi}{\psi},
\]
so that, using the measurement result $T=2.7255 \pm 0.0006$ K,
\[
\frac{|\delta \psi|}{\psi}\leq 2 (0.0006/2.7255) \simeq 4 \cdot 10^{-4},
\]
which improves Ni's bound on the dilaton field gradient by a factor of two. While such a light amplification factor would seem to indicate that it could also account for deviations from the standard Etherington distance duality relation even in the geometrical optics limit, we shall now see that this is, in fact, not the case.

\subsection{Conservation law}
The period-averaged energy-momentum tensor density (\ref{t-av}) can be interpreted physically in terms of a number current $N^i$ and momentum of light proportional to $k_j$ such that (e.g., \cite{Schneider:1992}, p.99)
\begin{equation}
\langle T \rangle^i{}_j=\sqrt{-g} N^i k_j.
\label{t-n}
\end{equation}
Now consider a bundle of light rays contained in the domain $\mathcal{D}$ in spacetime which is bounded by the intersection $\partial \mathcal{D}_O$ of the ray bundle with the observer's spacelike 3-space, the analogous boundary $\partial \mathcal{D}_S$ at the light source, and a null boundary traced out by light rays connecting  $\partial \mathcal{D}_S$ and  $\partial \mathcal{D}_O$. Then comparing the number observed at $O$ and emitted at $S$, one can define an excess fraction $\Delta$ of the total number $N$,
\[
\Delta\cdot N=\int_{\partial \mathcal{D}_O}\rmd^3x  \sqrt{h}\  N^i n_i -\int_{\partial \mathcal{D}_S}\rmd^3x \sqrt{h} \ N^i (-n_i), 
\]
where $h$ denotes the determinant of the Riemannian metric induced on the boundary by the Lorentzian spacetime metric $g_{ab}$, and $n_i$ is the outward pointing normal covector field. But since the integrand vanishes on the null part of the entire boundary $\partial D$ of $D$, we can recast the excess number as
\begin{equation}
\Delta\cdot N=\int_{\partial \mathcal{D}}\rmd^3x \sqrt{h} \ N^a n_a =\int_{\mathcal{D}} \rmd^4x \sqrt{-g}\  N^a{}_{;a} 
\label{del-n}
\end{equation}
where the last equality follows from Stokes' theorem. Clearly, $\Delta=0$ precisely if light rays are conserved, and its importance for the Etherington distance duality relation will be seen presently. Since, for the constitutive tensor density (\ref{chi}), light ray geometry is subject to the metric $g_{ab}$, the standard derivation of the Etherington distance duality relation applies, whence the flux seen by the observer $O$ from a light source $S$ of bolometric luminosity $L$ can be expressed as (cf. \cite{Schneider:1992}, p.115f)
\begin{equation}
S=\frac{(1+\Delta)L}{4\pi (1+z)^2 D_{\rm M}^2},
\label{eth1}
\end{equation}
where the distance $D_{\rm M}$ (also called transverse comoving distance,
\citep{Hogg:1999}) is related to the angular diameter distance $\da$ by
\begin{equation}
D_{\rm M}=(1+z)\da,
\label{eth2}
\end{equation}
which are both purely ray-geometrical and hence only $g_{ab}$-dependent. On the other hand, the luminosity distance $\dl$ is defined by connecting $L$ with the observable flux according to their Euclidean relationship,
\begin{equation}
S=\frac{L}{4\pi \dl^2}.
\label{eth3}
\end{equation}
Thus, combining (\ref{eth1}), (\ref{eth2}) and (\ref{eth3}), we derive a generalized Etherington distance duality relation,
\begin{equation}
\dl=\frac{(1+z)^2\da}{\sqrt{1+\Delta}},
\label{eth4}
\end{equation}
where $\Delta$ remains to be determined for our modified vacuum electrodynamics. Let us, then, take a step back to the energy-momentum tensor density defined by (\ref{t-gm}). The general Gotay-Marsden construction implies that the energy-momentum satisfies the conservation law 
\begin{equation}
T^i{}_{j,i}-T_{ab}g^{ab}{}_{,j}=0,
\label{con1}
\end{equation}
and applying the period-average of (\ref{con1}) to (\ref{t-av}), (\ref{t-n}) yields
\begin{equation}
(\sqrt{-g}N)^a{}_{,a}=0.
\label{con2}
\end{equation}
Therefore, it emerges that
\begin{eqnarray}
\sqrt{-g} N^a{}_{;a}&=& (\sqrt{-g}N)^a{}_{,a}+\left(\sqrt{-g}\Gamma^b{}_{ba}-\partial_a \sqrt{-g}\right)N^a \nonumber \\
\mbox{} &=& 0,
\label{n}
\end{eqnarray}
because on the right-hand side of the first equation, the first term vanishes on account of the conservation law (\ref{con2}) and the second term vanishes for the metric geometry defined by $g_{ab}$. Hence, using (\ref{n}) in (\ref{del-n}), we see that, indeed, $\Delta=0$ so that the standard Etherington distance duality relation (\ref{eth}) is recovered from (\ref{eth4}), even for this modified spacetime kinematics defined by the constitutive tensor density (\ref{chi}).
This shows that a non-zero $\Delta$ can only be obtained for a refined background geometry that locally violates Lorentz symmetry.

\section{Observational test of modifications}
\label{sec-obs}

In the standard cosmological model, the acoustic modes in the baryon photon
plasma at recombination imprint a distinct feature in the clustering amplitude
of galaxies \citep{Peebles:1970, Eisenstein:2005}. This feature,
called the Baryon Acoustic Feature (BAF hereafter), moves very little in
comoving coordinates and can be used as a standard ruler, whose physical value
can be calibrated based on observations of the Cosmic Microwave Background
fluctuations. Angular size measurements of the BAF can be used to infer the
angular diameter distance-redshift relation. The Sloan Digital Sky Survey III
Baryon Oscillation Spectroscopic Survey (SDSS BOSS Data Release 12) has
measured the redshifts and angular coordinates of nearly million galaxies at
redshifts $z\in[0.2, 0.7]$. Based on measurements of the BAF in two independent
redshift bins, constructed to have the same comoving volume, the collaboration
has posted the most precise measurements of the transverse comoving distances
\citep[see][for definition]{Hogg:1999}, $D_{\rm M}$ to effective redshifts of $0.38$ and
$0.61$\footnote{There is an intermediate bin which shares half of its volume
with the two bins used here.  We will use only the independent bins to avoid
use of the same data twice.}. These transverse comoving distances are related
to the angular diameter distance according to Eq.\ref{eth2}.

A variety of methodologies were developed within the BOSS collaboration to
extract the BAF and were subject to stringent tests using mock catalogs created
from simulated Universes in order to assess confidence in results
\citep{Alam:2016}. We use the constraints arrived at by consensus between
different methodologies as presented in \citep{Alam:2016}. These constraints
should be unaffected by any light extinction (or generation) events, as they
depend upon the angular size of the BAF. The values of the transverse comoving
distances to $z=0.38$ and $0.61$ have been determined with an accuracy of
$1.47$ and $1.39$ percent, respectively, with a cross-correlation coefficient
between the two measurement errors being about $20$ percent.

To obtain the observed luminosity distances to these redshifts, we make use of
the latest Joint Light-Curve Analysis sample of Type Ia supernovae from
\citet[][,B14 hereafter]{Betoule:2014}. The data set is a combination of
several SNe at low redshift ($z<0.1$), combined with data from all three
seasons of SDSS-II ($0.05<z<0.4$), and the Supernova Legacy Survey
($0.2<z<1.0$).  B14 perform intercalibration of the various surveys to reduce
the systematic uncertainties associated with calibration. They report the
observed redshifts (both heliocentric and cosmological), magnitudes and colors
at the peak, and the time-stretch factors of 740 Type Ia SNe
\citep{Betoule:2014}. The data set also includes the corresponding error
covariance matrix, which accounts for both the statistical and systematic
uncertainties in the data. B14 also provide software to compute the likelihood
of a set of luminosity distances predicted at the redshift of the SNe given the
data and the error covariance matrix.

We now describe constraints on deviations from the Etherington relation using
the latest compilation of luminosity distances and angular diameter distance
measurements. To remove any uncertainties in overall scaling of the individual
measurements (which can arise from uncertainties in the knowledge of the exact luminosity of the standard candle or the exact size of the standard ruler), and to obtain a test of the Etherington relation as independent of the cosmological model as possible, we follow the approach outlined in \cite{More:2009}, and test for the consistency of
\begin{equation}
\frac{D_{\rm L}(z_2)}{D_{\rm L}(z_1)} = \frac{D_{\rm A}(z_2)}{D_{\rm A}(z_1)} \frac{(1+z_2)^2}{(1+z_1)^2}\,.
\end{equation}
We will express any deviations from the above relation in terms of an opacity,
$\Delta \tau$ affecting the luminosity distances,
\begin{equation}
\frac{D_{\rm L, obs}(z_2)}{D_{\rm L, obs}(z_1)}e^{-\Delta\tau/2} = \frac{D_{\rm A}(z_2)}{D_{\rm A}(z_1)} \frac{(1+z_2)^2}{(1+z_1)^2}.
\end{equation}
Defining the distance modulus $\mu=25+5.0\log{D_{\rm L, obs}/{\rm Mpc}}$, one obtains
\begin{equation}
\Delta \tau = \frac{\ln (10)}{2.5} \Delta \mu(z_2, z_1) - 2.0 \ln\left[ \frac{(1+z_2)^2}{(1+z_1)^2} \frac{D_{\rm A}(z_2)}{D_{\rm A}(z_1)} \right]\ .
\end{equation}

To determine the observed difference in the distance moduli between effective
redshifts of $0.38$ and $0.61$ given the Supernova data, one has to make a choice
of the redshift bin from which we choose the SN. The SN within the particular
bin are then fit with a simple parametric model (e.g., a constant or a linear
relation), to determine the luminosity distances at the effective redshifts
\citep[see e.g.][]{More:2009}. This choice of the bin width then determines the
accuracy of these determinations. A small bin width, or using the nearest SN to
the determination of the angular diameter distance as done in
\cite{Holanda:2010, Liao:2015}, will translate into a small number of SNe
falling in to the bin, resulting in an increased sensitivity to the unmodelled
scatter in the apparent brightness of SN. A large bin width, on the other hand, 
raises the possibility that the parametric model may not be a good description
of the data, and thus introduces biases in the inferred luminosity distances.

The same problem arises while extracting the angular diameter
distance measurements from the BAF as well. The galaxies used to measure the
BAF also span a wide redshift range. Thus, in the case of the BAF, a fiducial
cosmological model is used to convert the observed angular and redshift space
separations of galaxies into separation in comoving coordinates. The location
of the BAF is measured using these converted coordinates. If the fiducial model
is indeed a good description of the data, then the predicted
location of the BAF and the observed one should match. A deviation constrains
the ratio of the angular diameter distance as inferred from the observations to
that of the angular diameter distance in the cosmological model.

\begin{figure*}[t]
\centering
\includegraphics[scale=1.2]{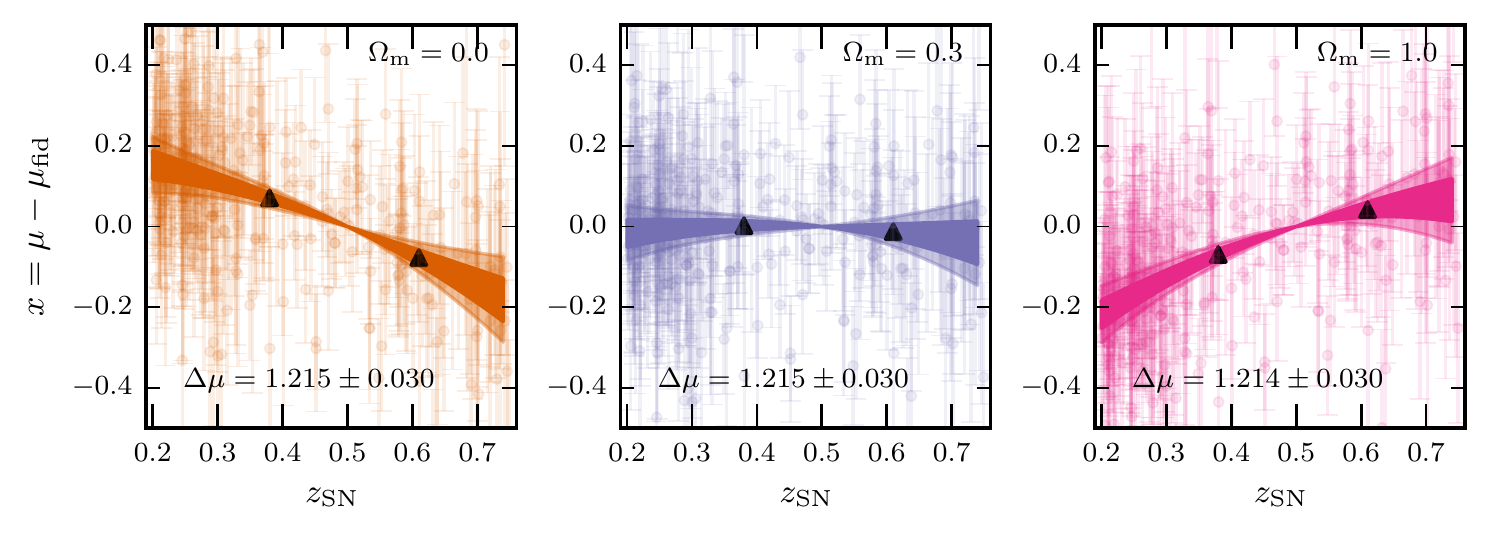}
\caption{Determination of $\Delta\mu$ from the SN data: The symbols with
errorbars in each of the three panels show the observed distance modulus $\mu$
minus the distance modulus in a fiducial cosmological model as indicated in the
top. The shaded bands indicate the 68 and 95 percent confidence regions from
our parametric fit to the data. The triangles indicate the redshifts at which
the BAO angular diameter distances were measured. The inferred value of
$\Delta\mu$ between these redshifts is indicated in the bottom of each panel.
The choice of the fiducial cosmological model has very little bearing on the
inferred value of $\Delta\mu$.}
\label{fig1}
\end{figure*}

We devise a new approach to analyze the SNe data, which treats the SN and BAO
data in a consistent fashion. We use all $337$ supernovae within the redshift
range $[0.2, 0.7]$, the same redshift range used for the measurement of the
BAF. We modelled the distance moduli to these SNe as the distance modulus in a
fiducial cosmological model, $\mu_{\rm fid}(z)$ plus up to a quadratic
correction in redshift,\footnote{This code along with a python wrapper for the
JLA likelihood code is available at http://github.com/surhudm/jla\_python .}
\begin{equation}
\mu(z) = \mu_{\rm fid}(z) + a_1 (z-z_{\rm anchor}) + a_2 (z-z_{\rm anchor})^2 \ ,
\label{eq:mucorr}
\end{equation}
where $a_1$ and $a_2$ are free parameters that encode the deviation of the distance
modulus from the fiducial model we use, and $z_{\rm anchor}=0.5$. We
have tested that changing $z_{\rm anchor}$ to 0.38 or $0.61$ gives identical
results. For the fiducial cosmological model, we use the flat $\Lambda$CDM
cosmological model with the matter density parameter $\Omega_{\rm m}=0.3$. 
The JLA likelihood model requires 4 additional nuisance parameters which encode
the absolute magnitude of SneIa, the stellar mass dependent correction to it,
and the stretch factor and color corrections, respectively. Note that we do not
include a constant term in Eq.~\ref{eq:mucorr} as such a term is exactly degenerate with the
nuisance parameter related to the absolute magnitude of SNeIa, and cancels out
when taking differences between the distance moduli at two different redshifts,
as we intend to do. The addition of the distance modulus from a fiducial
cosmological model reduces the non-linearities in the $\mu(z)$ relation. We
used the Monte Carlo Markov chain technique to sample the posterior
distribution of the parameters $a_1$ and $a_2$ given all the SNe with
$z\in[0.2, 0.7]$ after marginalizing over all the nuisance parameters. We then
used these samples of $a_1$ and $a_2$ to infer the difference in the distance
moduli at $z=0.61$ and $0.38$. We obtain $\Delta \mu = 1.215\pm0.030$ (see
middle panel of Fig.~\ref{fig1}). The fiducial cosmological model we have used
to derive this result has very little bearing on the result other than
providing a convenient way to reduce the non-linearities in the luminosity
distance redshift relation, which may be confirmed as follows: even using a flat 
$\Lambda$CDM model with $\Omega_{\rm m}=1.0$ or $\Omega_{\rm m}=0.0$, which is 
very much inconsistent with the data without the correction, causes a change to the 
inferred $\Delta\mu$ which is significantly smaller than the quoted error (compare the
left and the right panels of Fig.~\ref{fig1} with the middle panel).

Using these measurements of the angular diameter and luminosity distances, we
infer $\Delta \tau = \tau(0.61) - \tau(0.38) = -0.006 \pm 0.046$ (68 percent
confidence), perfectly consistent with our null hypothesis of no statistically
significant deviation from the Etherington relation. This extends the results
of \citet{More:2009} who measured $\Delta \tau$ between $0.2$ and $0.35$ to
higher redshifts and at greater precision (errors better by more than factor
2). One of the major improvements since the analysis of \citet{More:2009} has
been the separate measurement of the BAF in the transverse and the radial
direction, allowing for a direct measurement of the angular diameter distance.
This has allowed us to perform a more robust test of the Etherington relation
without having to marginalize over the cosmological model.

The current error budget on $\Delta \tau$ has a contribution of $0.028$
from the supernovae observations and $0.036$ from the BAO measurements (which results
in a total error on $\Delta\tau$ of $0.046$, when added in quadrature). Deeper
surveys such as the Subaru Hyper Suprime-Cam Survey \citep{Miyazaki:2012} may
be able to extend the redshift range even further by discovering SN at higher
redshifts, thus allowing us to test consistency with the angular diameter
distances measured at redshifts greater than two from the Lyman alpha forest
\citep{Busca:2013, Slosar:2013}.

\section{Conclusions}
\label{sec-con}
Observational advances are allowing stringent tests of the standard Etherington
distance duality relation. Although no statistically significant evidence for
deviations from the standard Etherington distance duality relation have been found
so far, theoretical studies of its possible modifications are needed to better
interpret observational data and elucidate the implications of deviations for
the fundamental kinematical structure of spacetime. 
\\ 
In this paper, we have considered, from a classical point of view, a minimal deviation 
from the standard kinematics of spacetime, given by the most general linear vacuum electrodynamics without birefringence, which introduces additional scalar fields while light rays are found to still be given by null geodesics of a Lorentzian spacetime metic. This kinematical structure of spacetime is defined by the constitutive tensor density (\ref{chi}). While it has been claimed that such a structure yields observationally accessible effects, we have shown that the Etherington distance duality relation remains unchanged.
\\
This suggests that, if mundane astrophysical sources such as dust can be ruled out, then a more refined kinematical stucture in which light rays are
not governed by a metric is required for fundamental modifications of the Etherington
relation. Since our analysis in section \ref{sec-th} has paid particular
attention to the conceptual basis of the Etherington relation, it becomes apparent 
how the non-metricity of a more general constitutive tensor density would affect the Etherington relation: although a generalized conservation law can be obtained also for a non-metrical spacetime structure according to Gotay-Marsden, a non-metric density in (\ref{n}) may still give rise to a non-zero excess $\Delta$ in (\ref{eth4}). In fact, a fully derived
example of this has recently been constructed \cite{Schuller:2016}, using a kinematical structure defined by an area metric tensor $G^{abcd}$ (e.g., \cite{Raetzel:2011}), whose gravitational dynamics for a point mass can be derived by means of the constructive gravity framework \cite{Schulleretal:2016} in the perturbative regime \cite{Schulleretal:2017}. 

Finally, we have also used the latest compilations of luminosity distance
determinations from SNeIa surveys as well as the angular diameter distance
determinations from measurements of the baryon acoustic feature, to perform an almost 
model independent and fairly robust test of the Etherington relation between
redshifts of $0.38$ and $0.61$. We find no statistically significant deviation
from the Etherington relation, and constrain the opacity of the Universe
between these two redshifts to be $\Delta\tau=-0.006\pm 0.046$.

\begin{acknowledgments}
We thank David Hogg, Shinji Mukohyama, Remya Nair, Tomomi Sunayama and Jean-Philippe Uzan for helpful discussions. 
\end{acknowledgments}

\bibliographystyle{apsrev}
\bibliography{sf}

\end{document}